# The Shifting Sands of Creative Thinking: Connections to Dual Process Theory

Paul Sowden and Andrew Pringle
University of Surrey

and Liane Gabora
University of British Columbia

For Correspondence: Paul Sowden, School of Psychology, University of Surrey, Guildford, GU2 7XH, UK. p.sowden@surrey.ac.uk

**Author Contributions**
Paul Sowden and Andrew Pringle wrote the first draft. Liane Gabora contributed critical review and additions to the manuscript. Paul Sowden produced the final version of the submitted manuscript.

**Word count main text:** 7,474

## Abstract
Dual process models of cognition suggest there are two kinds of thought: rapid, automatic *Type 1* processes, and effortful, controlled *Type 2* processes. Models of creative thinking also distinguish between two sets of processes: those involved in the generation of ideas, and those involved with their refinement, evaluation and/or selection. Here we review dual process models in both these literatures and delineate the similarities and differences. Both generative and evaluative creative processing modes involve elements that have been attributed to each of the dual processes of cognition. We explore the notion that creative thinking may rest upon the nature of a shifting process between generative and evaluative modes of thought. We suggest that through a synthesis application of the evidence bases on dual process models of cognition and from neuroimaging, together with developing chronometric approaches to explore the shifting process, could assist the development of interventions to facilitate creativity.

**Keywords:** creativity, dual process model, associative thought, analytic thought, Type 1 and Type 2 thinking

## Dual Process Models of Cognition

The notion that there are different types of thought processes extends at least back to the Ancient Greeks who distinguished between intellectual intuition and knowledge received by the senses. The nature of intuition and the distinction between sensory knowledge and reasoning also preoccupied later philosophers such as Descartes, Locke, Kant, Bergeson, and Russell for centuries (see Frankish, 2010). In the last decade, Cognitive Psychology has provided increasing evidence of two kinds of thinking processes and has made significant progress towards understanding their nature and operation (Evans, 2008; Frankish, 2010). One kind of process, often grouped as 'System 1' are characterised as rapid, unconscious, automatic, and associative in nature, corresponding to our gut reactions and intuitions. This system is seen as high capacity, able to rapidly combine (or associate) information that has been implicitly stored in memory, over long time periods, with sensory information from the current context without effortful thinking and intervention; that is, it is reflexive. The other set of processes are often grouped as 'System 2'. These involve thinking that is characterised as slow, controlled, effortful, conscious, and analytic, and that is limited by the capacity of working memory. Thinking using these processes is seen as rule based, applying these rules explicitly to current information; that is, it is reflective. Whereas System 1 processes are thought to be unrelated to intellectual ability, System 2 processes are thought to be correlated with intelligence. In some views, those with higher cognitive ability tend to be better able to apply System 2 thinking to intervene with belief based conclusions drawn from System 1 (Stanovich, 1999).

However, the simplicity of the distinction between these two systems drawn here belies the diversity and complexity of research in this area, and it is clear that the two systems account needs to be elaborated and revised. For instance, Evans (2008) notes that the evidence does not support the idea of just two separate systems, and he argues that we should stick with the terms Type 1 & 2 processes. He points out that there are different, incompatible variants of System 1, with those focusing on heuristic thinking quite different from those focused on associative thinking. Further, he notes that there are multiple forms of implicit processing that may be involved in 'System 1', some of which act autonomously to directly influence behaviour, and some of which supply pre-attentive information to working memory for analytic System 2 processes to use to control behaviour. In a related vein, Stanovich (2004) proposes The Autonomous Set of Systems (TASS), which have in common their fast, automatic, unconscious nature, but that differ in their diverse foci, which include emotional regulation of behaviour, unconscious implicit learning, and decision making principles that have become automatic through practice. Finally, in agreement with the view that System 1 might be better considered in terms of the set of underlying processes, Glöckner and Witteman (2010) proposed that intuition, a key feature of 'System 1', can in fact be broken down into four distinct kinds of automatic process. In sum, the evidence suggests that the notion of a unitary System 1 is not tenable.

A further issue is how the two types of processes interact. According to what Evans (2008) terms a "default interventionist" theory, Type 1 processes provide fast, preconscious information to working memory that is used to guide behaviour by default, unless analytic reasoning intervenes (e.g. Evans, 2006; Kahneman & Frederick, 2002). According to what Evans terms a "parallel-competitive" theory, implicit knowledge from Type 1 processes and explicit knowledge from Type 2 processes compete to control behaviour (e.g., Sloman, 1996; Smith & DeCoster, 2000). It has also been suggested that a third process controls the shift between Type 1 and Type 2 thinking (Stanovich, 2009). Use of the two systems may vary as a function of disposition such that some individuals are more inclined to use Type 1 processes and others Type 2 processes (e.g., Epstein, 2003; Stanovich, 1999). We contend that the nature of the interaction between the systems may be a key source of individual differences in creative thinking ability and of variation in creative thinking across time by a given individual. Before we can consider how this interaction might influence creative thinking it is necessary to consider the relationship between dual process models of cognition in general and dual process models of creativity more specifically.

## Dual Processes in Creative Thinking

Contemporary definitions of creativity vary considerably. Kaufman and Sternberg (2010, p. *xiii*) focus on the creative outcome and suggest that a creative response must be "novel, good, and relevant". Some add that creative responses should also be surprising (Boden, 2004; Simonton, 2012). Other definitions of creativity put less focus on the creative outcome. Plucker, Beghetto, and Dow (2004) suggest that "*creativity is the interaction among aptitude, process and environment by which an individual or group produces a perceptible product that is both novel and useful as defined within a social context*". Thus, their definition acknowledges that creativity arises in response to a confluence of factors both within and external to the individual (Csikszentmihalyi, 1996). Gabora (2010a) puts even more emphasis internal rather than external change, claiming that a cognitive process is viewed as creative to the extent that it transforms how the individual makes sense of or interacts with the world: "*The end result is that one's internal web of understandings is forged in a unique way, which is in turn reflected in its creative output*" (Gabora. 2010a). Despite these varying views, most contemporary researchers agree that creative <u>outcomes</u> must be both original and of value within a given domain. We use the term 'value' here because we believe it is more domain neutral than other terms used to encapsulate the quality aspect of creative responses, such as 'useful' or 'relevant', that may be inappropriate for domains such as the fine arts.

The assumption underlying several measures of creativity appears to be that creativity is the product of a single type of process, divergent thinking (e.g. the Alternate Uses Test and the Torrance Tests of Creative Thinking). However, creativity can also benefit from convergent thinking (Dietrich, 2007; Gabora, 2000, 2003), and evaluation (e.g. Silvia, 2008). A attempt to locate creativity in just one type of thinking process comes from research within the framework of Epstein (2003) proposed a Cognitive-Experiential Self Theory (CEST) of personality in which there are two information-processing systems: an experiential system that can be mapped onto Type 1 processes, and a rational system that can be mapped onto Type 2 processes. Epstein suggested that creativity is a product of the experiential system because, as an associative system, it can suggest ideas that would not be available to the linear-processing rational system. Indeed an experiential thinking style was found to be positively correlated with performance on divergent thinking tests (Norris & Epstein, 2011). However, these tests focus on fluency—how many creative ideas are generated—whilst neglecting comparative evaluation of the responses, nor do they require the kind of sustained critical creative thinking that is required by most real-world creative tasks.

The majority of other models of creative thinking tend to have in common an emphasis on the dual processes of developing ideas and evaluating them. The generic definition of a creative outcome implies these different processes, with generative processes required to formulate original ideas and evaluation processes required to select and/or refine those ideas into a form that is of value (see also Howard-Jones & Murray, 2003). We next review key models starting with the work of Guilford (1950, 1956), and examine their relationship to each other and to more general dual process models of cognition. We will see that while dual process models of creativity frequently appeal to the language of dual process models of cognition (e.g. Gabora & Aerts, 2009; Howard-Jones & Murray, 2003; Kaufman, 2011) they cannot always be straightforwardly mapped onto the dual processes espoused in models of cognition.

In his Structure of Intellect Model, Guilford (1956) distinguished between *divergent* and *convergent* thinking processes and the operation of these two processes can be viewed as an initial, contemporary dual process model of creative thinking. As mentioned above, divergent processes are traditionally thought to involve generating many different possible ideas, whereas convergent processes home in on a single solution. Several authors have suggested that divergent processes are associative in character, such that items encoded in memory are combined with information from the current context in a state of defocused attention (Gabora, 2010a; Martindale, 1999). Conversely, convergent processes are seen as analytic in character, and thought to predominate in the refinement and evaluating of solutions . Thus, in some respects divergent and convergent thinking appear to map neatly onto ideas about Type 1 and Type 2 processes. However, there are important differences. For instance, although divergent thinking may be spontaneous and associative in nature, with solutions sometimes appearing in a flash of insight, it can also be effortful and deliberate (e.g., Ward, 1994) and in this sense have more in common with Type 2 than Type 1 processes (Frankish, 2010). Furthermore, experiencing a flash of insight indicates convergence on an idea or solution but this may arrive at the end of an unconscious process (see ideas about incubation, Wallas, 1926) implying that convergent thinking can be unconscious and arise from Type 1 as well as Type 2 processes. Thus, a simple mapping of divergent thinking onto Type 1 and convergent thinking onto Type 2 processes is not possible.

Basadur *et al.* (1982; see also Basadur, 1995) propose the notion of ideation-evaluation cycles. They distinguish between three major stages in the creative thinking process — problem finding, problem solving, and solution implementation — and suggest that ideation and evaluation are involved at each stage in varying degree according to the domain. For instance, domains that emphasise problem finding have a high ratio of ideation to evaluation, whereas domains that emphasise solution implementation show the converse. They propose that dispositional differences in the tendency to ideate *vs*. evaluate lead to differences in the domain best suited to an individual.

An early dual process account of creativity is the *Darwinian theory of creativity,* sometimes referred to as BVSR for "Blind Variation and Selective Retention" (Campbell 1960; Simonton, 1999, 2007, 2010, 2012). It proposes we generate new ideas through an essentially a trial-and-error process involving *'blind'* generation of ideational *variants* followed by *selective retention* of the fittest variants for development into a final product. The variants are referred to as 'blind' because the creator has no subjective certainty about whether they are a step in the direction of the final creative result.

There is an intuitive appeal to BVSR; as do biological species, creative ideas exhibit the kind of complexity and adaptation over time that is indicative of an evolutionary process, both when they are expressed to others, and in the mind of a single creator (Gabora, 1997; Terrell, Hunt & Gosden, 1997; Thagard, 1980; Tomasello 1996). Although it is outside the scope of this paper to discuss them in detail, numerous problems have been identified with BVSR as a conceptual framework for creativity (Dasgupta, 2004; Eysenck, 1995; Gabora, 2005, 2007, 2010b, 2011; Sternberg, 1998; Weisberg, 2004). Nevertheless it seems reasonable that in creativity, as in natural selection, there is a process conducive to generating variety, and another conducive to pruning out inferior variants.

An assumption in BVSR is that successively generated ideas are unrelated. This is not supported by empirical studies of creativity (Feinstein, 2006; Nijstad & Stroebe, 2006). A dual process theory of creativity that is consistent with these findings is Finke, Ward and Smith's (1992) Genoplore model, which divides creative thinking into two overarching stages of idea generation and exploration. These are further subdivided into smaller stages with multiple operations available at each stage. For instance, generation can involve retrieval of items from memory, formation of associations between items, and synthesis and transformation of the resultant 'pre-inventive' structures. Exploration can involve identifying the attributes of these pre-inventive structures and considering their potential function in different contexts. Evidence for this model comes from findings that when people generate ideas they appear to make use of exemplars from the same or a related domain and they endow the new idea with many of the attributes of the previous exemplar. For instance, when asked to imagine an alien creature that lives on a planet very different from earth people typically include features of animals on earth such as arms and/or legs (Ward, 1994) Ward refers to this process as *the path of least resistance* (see also Ward & Kolomyts, 2010).

The Genoplore model can be partially mapped onto dual process models of cognition, with the generative phase corresponding to Type 1 thinking and the exploration of pre-inventive structures corresponding to Type 2 thinking. However, analytic processes may give rise to new ideas and insights (Finke, 1996). Thus, the generation phase appears to involve both Type 1 and 2 processes whereas the exploration phase is consistent with Type 2 processes alone.

The 'Honing Theory' of creativity (Gabora, 2005, in prep.) also posits that people draw on past knowledge when generating new possibilities, expanding on Mednick's (1962) work on flat associative hierarchies. Mednick found that whereas most people have rather steep associative hierarchies (i.e., a given stimulus evokes only highly related items in memory), creative people have flat associative hierarchies (i.e., a given stimulus evokes not just highly related but also remotely related items). Gabora proposes that creative people spontaneously enter a state of flattened associative hierarchies during idea generation (Gabora 2010a; see also Martindale, 1995), and that refinement of an idea occurs through iterative interaction between the current conception of the idea and the individual's internal model of the world, or 'worldview'. Creativity then arises through the joint effects of associative and analytic modes of thought and the process of shifting between them in response to task demands (for related views see also Howard-Jones, 2002; Martindale, 1999; Vartanian, Martindale & Kwiatkowski, 2007). In this view, the activation of flattened associative hierarchies, when presented with a task requiring a creative response, facilitates the forging of connections between more remote attributes of ideas and concepts in a contextually bound way that contrasts with the BVSR view that creative generation is independent of context. The latter view is necessary for BVSR because if generated ideas were influenced by context then they would likely be related to that context, and therefore to each other, and therefore not blind. The ideas that result from the associative process described by Gabora are then honed by an analytic process characterised by 'spiky' neural activation functions. Thinking becomes more focused on core aspects of the idea and can involve logically testing and elaborating the idea into a final solution.

Gabora's theory of creativity addresses criticisms of BVSR by accounting for observed effects of context and expertise on creative thinking (e.g., Howard-Jones & Murray, 2003; Vincent, Decker & Mumford, 2002) and gets us closer to an understanding of the fundamental mechanisms underlying creative thought, but there are still aspects to work out. The theory emphasises the importance of the ability to shift from an analytic to an associative mode of thought when stuck in a rut, and from an associative to an analytic mode following insight, and computer models have shown this shifting to be effective (DiPaola & Gabora, 2007, 2009; Gabora, Chia & Firouzi, 2013; Gabora & DiPaola, 2012). However, there has been little empirical investigation of this shifting in humans. Thus, when analysed in relation to dual process models of cognition, it is not yet clear whether a straightforward mapping can be made.

Whilst associative and analytic thinking processes appear to be closely related to Type 1 and 2 thinking, their recruitment may be different. In dual process models of cognition, Type 1 processes either govern behaviour by default or in competition with Type 2 processes. However, the nature of the relationship between these types of process in creative thinking may be different. For example, creative thinking may be characterised by shifting from analytic to associative thinking as suggested in Howard-Jones (2002) Dual-State model of creative cognition. Howard-Jones suggests that a key barrier to creativity is the notion of 'fixation', and proposes strategies to help shift individuals from secondary process, analytical thinking to primary process associative thinking. At present, Gabora's theory is uncommitted to where on the spectrum from associative to analytic the 'default' starting point falls when thinking creatively and acknowledges the possibility that the default point may vary across individuals (see also Basadur *et al.* 1982; Basadur, 1995; Epstein, 2003; Stanovich, 1999). It seems likely that work on dual process models of cognition could usefully contribute to exploring the process of shifting between Type 1 and 2 thinking processes in the context of creative thinking.

Another recent dual process theory of creative thinking, proposed by Nijstad, De Dreu, Rietzschel and Baas (2010), suggests that creativity can arise through two pathways: a *flexibility* pathway and a *persistence* pathway. Greater cognitive flexibility is viewed as generating more categories of ideas, and as more frequent shifting between them, which Nijstad *et al.* argue can lead to greater originality. Thus, the flexibility pathway appears to relate to dual-process models of creativity, such as BVSR, that have suggested that multiple idea variants are generated. However, they argue that exploring a few content categories in depth may also lead to increased originality. Given its focus on a few categories of idea, the persistence pathway appears to relate to other models of creativity that focus on developing ideas, such as Honing Theory. However, the flexibility and persistence model is different from these previous models in that it strongly emphasises deliberative processing under conscious executive control. Indeed Nijstad *et al.* (p43) state "*our model is not applicable to situations in which creativity occurs ''spontaneously'' without intentional effort*". In essence, they argue that the creative thinking described by their model is primarily a product of Type 2 processes. However, they note that the degree of executive control differs between their two pathways. The flexibility pathway uses broad and inclusive cognitive categories, flexibly shift among categories, approaches, and sets, and establish more remote associations. The authors suggest that, in the case of individuals with high mental ability, this may be helped by reduced latent inhibition, which allows more distant associates and ideas to enter working memory. Such individuals may have the necessary cognitive control to benefit from reduced inhibition because they are also able to evaluate and identify the inevitable proportion of poor solutions that result if more are allowed. The persistence pathway achieves creative outcomes through a systematic and effortful exploration of possibilities and an in depth exploration of only a few categories or perspectives. These systematic processes are likely to start with readily available solutions. Less obvious, more original, solutions only arrive if the individual persists in generating more ideas within a category. Because systematic search is involved in this pathway they argue that higher levels of executive control are required than for the flexibility pathway. When persisting, individuals will be less distractible, but also less flexible because more distant associates are filtered out.

Nijstad *et al.* note that because the persistence pathway reduces flexibility it might appear to be negatively related to the flexibility pathway. However, they argue that this is not the case because individuals can shift from a flexible mode, where they discover new and promising approaches to a task, to a systematic, persistent mode where they further explore these approaches. In describing this shifting process the distinction from previous 'generate and explore' models is somewhat lost. It again appears that creativity may most likely result from the joint operation of both pathways such that the flexibility pathway helps with the development of originality and the persistence pathway with the elaboration of ideas into those that will prove useful. As such the pathways might not appear conceptually to be independent 'dual-routes' but necessarily coupled stages in a creative thinking process, much as proposed in other work. However, arguing against this, Nijstad *et al* (pers. com.) find that the correlation between their measure of flexibility – the number of categories generated – and their measure of persistence - the <u>within</u> category fluency (average number of ideas per category/total number of ideas generated) – decreases over time. This might suggest that on a given task, rather than shifting between pathways individuals become increasingly entrenched in the processing associated with one pathway over time.

Nijstad *et al.* provide a meta-analysis of the relationship between their measures of flexibility and persistence on the originality and feasibility of creative ideas. They find a correlation of 0.36 between flexibility and originality that is much stronger than the correlation of 0.12 between within-category-fluency (persistence) and originality. This suggests that the flexibility pathway is more strongly implicated in the overall originality of ideas, although Nijstad et al cite evidence that the relationship between within-category-fluency and originality increases for the originality of ideas within a particular category (Rietzschel, 2005; Rietzschel, Nijstad & Stroebe, 2007). This latter finding somewhat strengthens the argument that persistence can indeed increase originality, but still overall originality is much more strongly related to flexibility than to persistence. In addition, Nijstad *et al* find no factors that significantly correlate with feasibility of ideas other than originality, which is negatively correlated. This, coupled with a greater emphasis on measures of originality, leaves somewhat unaddressed in their model the question of how creative ideas that are both original <u>and of value</u> come about. One possibility arises from examination of the effects of mood state on creativity. De Dreu et al (2008) find that negative activating mood states enhance persistence, which they argue should benefit within category fluency. However, another possibility is that negative activating mood states also improve the evaluation of creative ideas as observed by Sowden & Dawson (2011) thereby enhancing their value.

In summary, Nijstad *et al* provide strong evidence for a flexibility pathway that promotes creativity through increasing the originality of ideas, but the evidence for the impact of a second persistence pathway on originality is weaker, nor does persistence relate to the feasibility (value) of ideas. The factors predicting the latter are largely unexplained by their model. Thus again, whilst dual process models of cognition clearly relate to Nijstad *et al'*s ideas, the mapping is not straightforward. There appears to be a clear relationship between Type 1 processes and the flexibility pathway in that both are associative in character, but the flexibility pathway requires deliberative processing to evaluate ideas ("*an idea monitor*") and keep them on track, sometimes viewed as a function of Type 2 processing. Further, although both Type 2 processes and the persistence pathway are hypothesised to make strong use of working memory, the persistence pathway does not appear to contribute to the feasibility of creative ideas whereas a key function of Type 2 processes is evaluation.

In summary, we have looked at two sets of dual process models: general cognitive models and models of creative thinking. Table 1 summarizes the various elements of these models and the relationships between them.

[Insert Table 1 about here]

It is clear that both idea generation and evaluation may recruit Type 1 and 2 processes, and this might suggest the possibility of a four-component model of creative thinking. Although this cannot be ruled out, some recent models of creative thinking do align quite parsimoniously with Dual Process models of cognition, particularly Howard Jones' Dual State model (2002) and Gabora's Honing Theory (2005, in prep.). Nevertheless, the question of how Type 1 and 2 processes interact in creative thinking remains largely unanswered. Although there is general agreement that shifting between the different modes is required for creative ideas to develop (e.g. Basadur, 1982; Basadur *et al.*, 1985; Finke *et al.*, 1992; Gabora & Ranjan, 2013; Howard-Jones 2002; Nijstad *et al.*, 2010), it is not known whether this is different when thinking creatively, how it varies across individuals, and whether it could be trained. These issues are addressed in the next sections.

**Mechanisms of shifting between modes of creative thought**

Dual-process models of creative thinking make different predictions concerning the extent to which the modes operate in series or in parallel. The dual-stage models of Basadur (1995), Howard-Jones (2002), Finke *et al.* (1992) and the theory of the emergence of a creative insight proposed by Gabora and Ranjan (2013) suggest shifting between modes of thinking occurs in series. However, Nijstad *et al's* (2010) dual-pathway model suggests that an 'idea monitor', a mechanism that evaluates ideas, continually checks the outputs of idea generation processes in a similar fashion to type 2 processes in default-interventionist dual-process accounts of cognition (Evans, 2008; Frankish, 2010). This suggests that individuals have the mode of thinking supporting evaluation available ''on tap' and can apply it to keep idea generation in check. While not necessarily implying that the two modes ] operate in parallel, Nijstad *et al.'s* (2010) model does suggest that generation and evaluation work more in unison than in the aforementioned serial models. Serial models can be interpreted as implying that it is necessary to disengage one mode prior to engaging the other mode, or shift along the continuum between analytic to associative thinking (Gabora & Ranjan, 2013). If shifting does occur in a serial fashion then attempts to stimulate it could focus on improving the individual's ability or tendency to engage the optimal mode of thinking for performance in a particular phase of the creative process. On the other hand if generation and evaluation occur in a more parallel fashion then it may be best to improve an individual's evaluative ability and ensure that evaluative processes are quickly accessible 'on tap' alongside associative processes. It could be that both parallel and serial accounts are correct but there are individual differences in the shifting mechanism. For example, creative individuals may have evaluative processes 'on tap' while less creative individuals may shift more slowly between different modes in a serial fashion.

Different models also propose different mechanisms that enable shifting between modes of thinking. Nijstad *et al.* (2010) propose that executive control underlies thinking within both pathways but lower cognitive control is applied in the flexibility pathway when broadening one's attention, while higher cognitive control is applied in the persistence pathway to support systematic search. With different pathways conceptualised as engaging different levels of cognitive control, a means of stimulating shifting between them may take the form of stimulating a person's ability to flexibly modulate cognitive control in response to the phase of the creative process. There is some evidence that creative people are better able to flexibly modulate cognitive control in response to context (Zabelina & Robinson, 2010). However, further work is needed to determine if training individuals to flexibly modulate cognitive control could improve their creativity.

Another mechanism that has been suggested to enable shifting between modes involves differentially adjusting the focus of attention based on the demands of the task at hand (Gabora, 2003; Vartanian, Martindale & Matthews, 2009). There may be a link between the mechanism responsible for adjustments in cognitive control and the mechanism responsible for adjusting attentional focus. For instance, Kaufman (2011) suggests that, during generative thinking, unconscious cognitive processes activated through defocused attention are more prevalent, whereas, during exploratory thinking, controlled cognition activated by focused attention becomes more prevalent. There is evidence that creative individuals are better able to alter this focus of attention in response to task demands by modulating cognitive inhibition (Ansburg & Hill, 2003; Bristol & Viskontas, 2006; Dorfman, Martindale, Gassimova & Vartanian, 2008; Gabora, 2000, 2003; Vartanian, Martindale & Kwiatkowski, 2007). In addition, shifting between these processes may be more effectively applied by creative individuals over the course of a creative task (Gilhooly, Fiortou, Anthony & Wynn, 2007; Vartanian, Martindale & Kwiatkowski, 2003). However, there is debate about whether shifting happens automatically or under top-down control (Vartanian, Martindale & Matthews, 2009), which would appear to hold implications for efforts to train this mechanism. If it is under top-down control then explicit instructions could be used to stimulate less creative individuals to better adjust their focus of attention, but if it occurs automatically then some means of implicitly inducing shifts in attentional focus may be required. It may be the case, as Vartanian, Martindale & Matthews (2009) suggest, that both top-down and automatic bottom-up processes can drive adjustments in the focus of attention, and that whether this mechanism is under conscious control, or not, is a function of the stage of the task that one is engaged in. Dual-process accounts of 'non-creative' cognition have suggested that a third level of processing, termed the reflective mind, might initiate the shift between Type 1 and Type 2 processes (Stanovich, 2009). These Dual-process theories could inform attempts to understand what appears at to be a mechanism that controls shifting between the two modes and hence could play a role in attempts to stimulate shifting.

New section heading? Moving to a neurobiological level of explanation, one potential candidate for a shifting mechanism is the 'salience network' proposed by Menon & Uddin (2010). The salience network incorporates the Anterior Insula and the Anterior Cingulate Cortex, which has been shown to become active shortly before an insightful solution is reached (Kounios & Beeman, 2009) consistent with it playing a role in the shift from unconscious generative processing to conscious validation. Menon & Uddin (2010) propose that this network serves to shift between a Default Mode Network (DMN) and a Central Executive Network (CEN; see also Raichle, MacLeod, Snyder, Powers, Gusnard & Shulman, 2001; Fox, Snyder, Vincent, Corbetta, Van Essen & Raichle, 2005). They summarise evidence that the DMN shows decreases in activity during cognitively demanding tasks and includes brain regions involved in self-referential and social-cognitive processing, and in processing episodic and autobiographical memory. It has been further suggested that this retrieval of information from memory, both personal and general may facilitate solving problems and developing future plans (Greicius, Krasnow, Reiss & Menon, 2003). Thus, the function of the DMN might appear to bear some relationship to the associative processes involved in the generation of creative ideas (see Buckner, Andrews-Hanna & Schacter, 2008). Conversely, CEN activity increases during cognitively demanding tasks and has been linked to maintaining and manipulating information in working memory and making judgements and decisions. Thus, the CEN may play a role in the analysis and evaluation of creative ideas and Type 2 thinking processes. Further, the CEN is closely aligned anatomically and functionally with a reflective C network proposed by Lieberman, Gaunt, Gilbert, and Trope (2002; see also Lieberman, Jarcho & Satpute, 2004) whilst the DMN shares some overlap in brain regions with a reflexive, X network proposed by Lieberman *et al*. (2002; 2004). However, the latter are not perfectly aligned anatomically and whether they are functionally aligned is not yet clear in the literature. The X-network is linked with associative learning that occurs without conscious intention and with intuition-based self-knowledge (Lieberman *et al*., 2002, 2004), which bears some relation to conceptions of the DMN. However, whereas episodic memory retrieval is seen as part of the DMN it is seen as a key aspect of the C network by Lieberman *et al.* (2004). Further illustrating the complexity of the relationship between generative and evaluative thinking, and the underlying brain mechanisms, a recent fMRI study by Ellamil, Dobson, Beeman & Krisstoff (2012) found that the default mode network and the central executive network were both activated during evaluation of creative ideas but not during creative generation. The latter was associated with activation of the medial temporal lobes, which supports memory retrieval during creative generation.

Clearly there is much to be worked out in the mapping between brain networks, the processes of creative thinking and dual process models of cognition. However, the consistent involvement of the ACC in shifting suggests that ACC activation could be a useful marker to identify shifting as participants work on creative problems. This could afford us a way to compare the timing and frequency of shifting between more and less creative problems solvers and solutions.

## Developing chronometric approaches to explore the time course of shifting

Methods for training creativity have tended to focus on just one component such as divergent thinking (Scott, Leritz & Mumford, 2004). However, given the evidence that creativity involves an interaction between divergent and convergent thinking (Basadur, 1995; Finke, 1996; Gabora & Ranjan, 2013; Howard-Jones 2002; Nijstad *et al.*, 2010), attempts to stimulate the shifting and interaction between these modes could be more effective.

Stimulating shifting between different modes of thinking to aid creativity could involve enhancing the means by which one shifts, the ability to shift, or the time course of shifting. An individual who is proficient at rapidly shifting between modes may not know *when* to shift. For instance, the finding that presenting examples during a generative phase can have detrimental effects (Jansson & Smith, 1991) suggests that a mismatch between the phase of the creative process and the mode of thinking can negatively affect creative output (Howard-Jones, 2002). Further, rapid cycling between modes of thinking may result in ideas being prematurely evaluated leading to promising ideas being dismissed (Nijstad, *et al.*, 2010; Zabelina & Beeman, 2013), whereas an individual who has a good understanding of when is the optimal time to shift might be better at allowing time for an idea to be sufficiently worked out before applying evaluative processes.

The possibility that patterns of shifting differ across stages of the creative process (Basadur, 1995) suggests that the relationship between creativity and shifting is an important avenue for further research. Factors such as frequency of shifts, the length of operating in one mode before a shift, and differences in patterns or the nature of shifting at different stages could be examined. Experimental studies using think-aloud protocols (Gilhooly *et al.*, 2007) in combination with neuroimaging methods such as EEG (Martindale & Hasenfus, 1978; Fink & Benedek, in press) open up the possibility of using chronometric approaches to explore the time course of shifting during creative thinking. By using them conjointly it may be possible to lock events in the creative process to patterns of neural activity exhibited by an individual while they are engaged in a creative act. This approach could be used to inform instruction to people of when it is optimal to shift in order to maximise creative output.

Finally, individual differences in trait dispositions to engage in different modes or styles of thinking have also been proposed (Basadur *et al.* 1982; Basadur, 1995; Epstein, 2003; Kaufman, 2011; Norris & Epstein, 2011; Stanovich, 1999). Chronometric approaches would allow us to compare whether these trait differences affect the timing of the shifting process. Differences in the tendency to engage in a certain mode of thinking could have implications for the design of interventions to stimulate shifting. For example, Howard-Jones (2002) argues that individuals evidencing a disposition to think more analytically may benefit from interventions which focus on enabling them to shift to a more associative mode of thinking while those evidencing a disposition to think more imaginatively may benefit from interventions that enable them to shift to a more self-critical, analytical style of thinking (Howard-Jones, 2002). Further researche is necessary to examine if trait differences interact with the process of shifting between modes of thinking.

## Summary and Conclusions

There appears to be consensus amongst different dual-process models of creativity on the importance of generative and evaluative processes and on the importance of the interaction between these modes during creative thinking. However different models appear to conceptualise this in different ways. The time seems ripe to develop an integrated dual-process model of creativity that clearly specifies the nature of this interaction across different points in the creative process and the mechanisms that underlie shifting between modes. An important part of this process will be to make use of the findings from the wider body of dual-process theories of cognition. By understanding the relationship between Type 1 and Type 2 thinking processes on the one hand and the processes of generation and evaluation, as conceived in different models of creativity, on the other, we may be better able to elaborate our understanding of the creative thinking process and its underlying biological mechanisms. This should help us to develop new methods of intervening to enhance creative thinking. In future studies it would be useful to investigate (1) the extent to which shifts in ideation-evaluation ratios can be achieved, and (2) whether creativity is enhanced by the ability to shift along the continuum as a function of the circumstances and the stage of the creative process. One particularly promising avenue is the use of a chronometric approach to explore patterns of shifting between modes whilst thinking creatively.

## References

Ansburg, P. I., & Hill, K. (2003). Creative and analytic thinkers differ in their use of attentional resources. *Personality and Individual Differences*, *34*, 1141–1152.

Basadur, M. S. (1995). Optimal ideation-evaluation ratios. *Creativity Research Journal*, *8*, 63-75.

Basadur, M. S., Graen, G., & Green, S. (1982). Training in creative problem solving: Effects on ideation and problem finding and solving in an industrial research organization. *Organizational Behavior and Human Performance*, *30*, 41-70.

Boden, M. A. (2004). *The creative mind: myths and mechanisms* (2nd ed.). New York, NY: Routledge.

Bristol, A. S., & Viskontas, I. V**.** (2006). Dynamic processes within associative memory stores: piecing together the neural basis of creative cognition. In J. C. Kaufman, & J. Baer (Eds) *Creativity, Knowledge and Reason*. Cambridge, UK: Cambridge University Press. pp. 60-80.

Buckner, R. L., Andrews-Hanna, J. R. & Schacter, D. L. (2008). The brain's default network. *Annals of the New York Academy of Sciences*, *1124*, 1-38.

Campbell, D. T. (1960). Blind variation and selective retention in creative thought as in other knowledge processes. *Psychological Review*, *67*, 380-400.

Csikszentmihalyi, M**.** (1996). *Creativity: flow and the psychology of discovery and invention*. New York, NY: Harper Perennial.

Dasgupta, S. (2004). Is creativity a Darwinian process? *Creativity Research Journal, 16,* 403−413.

De Dreu, C. K. W., Baas, M., & Nijstad, B. A. (2008). Hedonic tone and activation level in the mood-creativity link: Toward a dual pathway to creativity model. *Journal of Personality and Social Psychology*, *94*, 739-756.

DiPaola, S., & Gabora, L. (2007). Incorporating characteristics of human creativity into an evolutionary art algorithm. In D. Thierens (Ed.), *Proceedings of the Genetic and Evolutionary Computing Conference (GECCO)* July 7-11, University College London, England. pp. 2442-2449.

DiPaola, S., & Gabora, L. (2009). Incorporating characteristics of human creativity into an evolutionary art algorithm. *Genetic Programming and Evolvable Machines, 10*, 97-110.

Dietrich, A. (2007). Who's afraid of a cognitive neuroscience of creativity? *Methods*, *42*, 22-27.

Dorfman, L., Martindale, C., Gassimova, V., & Vartanian, O. (2008). Creativity and speed of information processing: A double dissociation involving elementary versus inhibitory cognitive tasks. *Personality and Individual Differences*, *44*, 1382-1390.

Ellamil, M., Dobson, C., Beeman, M., & Christoff, K. (2012). Evaluative and generative modes of thought during the creative process. *Neuroimage*, *59*, 1783-1794.

Epstein, S. (2003). Cognitive-experiential self-theory of personality. In Millon, T., & Lerner, M. J. (Ed's), *Comprehensive Handbook of Psychology, Volume 5: Personality and Social Psychology*. Hoboken, NJ: Wiley & Sons. pp. 159-184.

Evans, J. St. B. T. (2006). The heuristic-analytic theory of reasoning: extension and evaluation. *Psychonomic Bulletin and Review*, *13*, 378–95.

Evans, J. St. B. T. (2008). Dual-process accounts of reasoning, judgment and social cognition. *Annual Review of Psychology, 59,* 255-278.

Eysenck, H. J. (1995). *Genius: The natural history of creativity*. Cambridge, England: Cambridge University Press.

Feinstein, J. (2006). The nature of creative development. Stanford CA: Stanford University Press.

Fink, A., & Benedek, M. (in press). EEG alpha power and creative ideation. *Neuroscience & Biobehavioral reviews.*

Finke, R. A. (1996). Creative insight and preinventive forms. In R. J. Sternberg & J. B. Davidson (Ed's), *The Nature of Insight*. Cambridge, MA: MIT Press. pp. 255-280.

Finke, R. A., Ward, T. B., & Smith, S. M. (1992). *Creative Cognition: Theory, Research and Applications.* Cambridge, MA: MIT Press.

Fox M. D., Snyder A. Z., Vincent J. L., Corbetta M., Van Essen D. C., & Raichle M.E. (2005). The human brain is intrinsically organized into dynamic, anticorrelated functional networks. *Proceedings of the National Academy of Sciences of the USA*, *102*. 9673–9678.

Frankish, K. (2011). Dual-process and dual-system theories of reasoning. *Philosophy Compass, 10*, 914-926.

Gabora, L. (2000). Toward a theory of creative inklings. In R. Ascott (Ed.) *Art, Technology, and Consciousness*. Bristol, UK: Intellect Press. pp. 159-164.

Gabora, L. (2003). Contextual focus: A cognitive explanation for the cultural transition of the Middle/Upper Paleolithic. In R. Alterman & D. Hirsch (Ed's) *Proceedings of the 25th Annual Meeting of the Cognitive Science Society*, Boston MA, July 31-August 2. Hillsdale, NJ: Lawrence Erlbaum Associates. pp. 432-437.

Gabora, L. (2005). Creative thought as a non-Darwinian evolutionary process. *Journal of Creative Behavior, 39*, 262-283.

Gabora, L. (2010a). Revenge of the 'neurds': Characterizing creative thought in terms of the structure and dynamics of human memory. *Creativity Research Journal, 22*, 1-13.

Gabora, L. (2010b). Why Blind-Variation-Selective-Attention is inappropriate as an explanatory framework for creativity. *Physics of Life Reviews, 7*, 190-194.

Gabora, L. (2011). An analysis of the Blind Variation and Selective Retention (BVSR) theory of creativity. *Creativity Research Journal, 23*, 155-165.

Gabora, L., & Aerts, D. (2009). A model of the emergence and evolution of integrated worldviews. *Journal of Mathematical Psychology*, *53*, 434-451.

Gabora, L., Chia, W. W., & Firouzi, H. (2013). A computational model of two cognitive transitions underlying cultural evolution. *Proceedings of the 35th Annual Meeting of the Cognitive Science Society,* July 31 - Aug. 3, Berlin. Houston, TX: Cognitive Science Society. pp. 2344-2349.

Gabora, L., & DiPaola, S. (2012). How did humans become so creative? *Proceedings of the International Conference on Computational Creativity,*. May 31 - June 1, Dublin, Ireland. pp. 203-210.

Gabora, L. & Ranjan, A. (2013). How insight emerges in a distributed, content-addressable memory. In A. Bristol, O. Vartanian, & J. Kaufman (Ed's) *The Neuroscience of Creativity*. New York: Oxford University Press.

Greicius MD, Krasnow B, Reiss AL, & Menon V (2003) Functional connectivity in the resting brain: a network analysis of the default mode hypothesis. *Proceedings of the National Academy of Sciences of the USA*, *100*, 253–258.

Gilhooly, K., Fioratou, E., Anthony, S.H., & Wynn, V. (2007). Divergent thinking: Strategies and executive involvement in generating novel uses for familiar objects. *British Journal of Psychology, 98*, 529-694.

Glöckner, A., & Witteman, C. (2010). Beyond dual-process models: A categorisation of processes underlying intuitive judgement and decision making. *Thinking and Reasoning*, *16*, 1-25.

Guilford, J. P. (1950). Creativity. *American Psychologist*, *5*, 444-454.

Guilford, J. P. (1956). The structure of intellect. *Psychological Bulletin*, *53*, 267-293.

Howard-Jones, P. A. (2002). A dual-state model of creative cognition for supporting strategies that foster creativity in the classroom. *International Journal of Technology and Design Education, 12*, 215-226.
Howard-Jones, P.A., & Murray, S. (2003). Ideational productivity, focus of attention and context. *Creativity Research Journal*, *15*, 153-166.
Jansson, D. G., & Smith, S. M. (1991). Design fixation. *Design Studies, 12*, 3-11.
Kahneman D., & Frederick S. (2002). Representativeness revisited: attribute substitution in intuitive judgement. In T. Gilovich, D. Griffin and D. Kahneman (Ed's) *Heuristics and Biases: The Psychology of Intuitive Judgment*. Cambridge, UK: Cambridge University Press. pp. 49–81.
Kaufman, S. B. (2011). Intelligence and the cognitive unconscious. In R. J. Sternberg & S. B. Kaufman (Ed's), *The Cambridge Handbook of Intelligence*. Cambridge, UK: Cambridge University Press. pp. 442-467.
Kaufman, J. C., & Sternberg, R.J. (2010) *The Cambridge Handbook of Creativity*, New York, NY: Cambridge University Press.
Klein, G. (1999). *Sources of Power*. Cambridge, MA: MIT Press.
Kounios, J., & Jung-Beeman, M. (2009). Aha! The cognitive neuroscience of insight. *Current Directions in Psychological Science, 18,* 210-216.
Lieberman M. D., Jarcho J.M., & Satpute A.B. (2004). Evidence-based and intuition-based selfiknowledge: an fMRI study. *Journal of Personality and Social Psychology*, *87*. 421–435.
Lieberman, M. D., Gaunt, R., Gilbert, D. T., & Trope, Y. (2002). Reflection and reflexion: A social cognitive neuroscience approach to attributional inference. *Advances in Experimental Social Psychology*, *34*, 199-249.
Martindale, C. (1995). Creativity and connectionism. In S. M. Smith, T. B. Ward, & R. A. Finke (Ed's), *The Creative Cognition Approach*. Cambridge, MA: MIT Press. pp. 249–268.
Martindale, C. (1999). Biological bases of creativity. In R. J. Sternberg (Ed.), *Handbook of creativity* . New York: Cambridge University Press. pp. 137–152.
Martindale, C. & Hasenfus, N. (1978). EEG differences as a function of creativity, stage of the creative process, and effort to be original. *Biological Psychology, 6*, 157-167.
Menon, V., & Uddin, L. Q. (2010). Saliency, switching, attention and control: a network model of insula function. *Brain Structure and Function, 214*, 655-667.
Nijstad, B.A., De Dreu, C.K.W., Rietzschel, E.F. & Baas, M. (2010). The dual pathway to creativity model: Creative ideation as a function of flexibility and persistence. *European Review of Social Psychology, 21*, 34-77.
Nijstad, B. A., & Stroebe, W. (2006). How the group affects the mind: A cognitive model of idea generation in groups. *Personality and Social Psychology Review, 10,* 186–213.
Norris, P., & Epstein, S. (2011). An experiential thinking style: its facets and relations with objective and subjective criterion measures. *Journal of Personality*, *79*, 1043-1080.
Plucker, J. A., Beghetto, R. A., & Dow, G. T. (2004). Why isn't creativity more important to educational psychologists? Potentials, pitfalls, and future directions in creativity research. *Educational Psychology*, *39*, 83-96.
Raichle, M. E., MacLeod, A. M., Snyder, A. Z., Powers, W. J., Gusnard, D. A., & Shulman, G. L. (2001). A default mode of brain function. *Proceedings of the National Academy of Sciences*, *98*, 676–682.

Rietzschel, E. F. (2005). From quantity to quality: Cognitive, motivational, and social aspects of creative idea generation and selection. Unpublished doctoral dissertation, Utrecht University.

Rietzschel, E. F., Nijstad, B. A., & Stroebe, W. (2007). Relative accessibility of domain knowledge and creativity: The effects of knowledge activation on the quantity and quality of generated ideas. *Journal of Experimental Social Psychology, 43,* 933–946.

Scott, G., Leritz, L.E. & Mumford, M.D. (2004). The effectiveness of creativity training: A quantitative review. *Creativity Research Journal*, *16*, 361-388.

Silvia, P. J. (2008). Discernment and Creativity: How Well Can People Identify Their Most Creative Ideas? *Psychology of Aesthetics, Creativity and the Arts*, *2*, 139-146.

Simonton, D. K. (1999). Creativity as blind variation and selective retention: Is the creative process Darwinian? *Psychological Inquiry*, *10*, 309–328.

Simonton, D. K. (2007). The creative imagination in Picasso's Guernica sketches: Monotonic improvements or nonmonotonic variants? *Creativity Research Journal*, *19*, 329–344.

Simonton, D. K. (2010). Creative thought as blind-variation and selective retention: Combinatorial models of exceptional creativity. *Physics of Life Reviews*, *7*, 156-179.

Simonton, D. K. (2011). Creativity and discovery as blind variation: Campbell's (1960) BVSR model after the half century mark. *Review of General Psychology*, *15*, 158-174.

Simonton, D. K. (2012). Taking the U.S. Patent Office criteria seriously: a quantitative three-criterion creativity definition and its implications. *Creativity Research Journal*, *24*, 97-106.

Simonton, D. K. (in press). Creative Thought as Blind Variation and Selective Retention: Why Creativity is Inversely Related to Sightedness. *Journal of Theoretical and Philosophical Psychology*.

Sloman S. A. (1996). The empirical case for two systems of reasoning. *Psychological Bulletin*, *119*, 3–22.

Smith E.R. and DeCoster J. (2000). Dual-process models in social and cognitive psychology: conceptual integration and links to underlying memory systems. *Personality and Social Psychology Review*, *4*, 108–31.

Sowden, P. T., & Dawson, L. (2011). Creative feelings: the effect of mood on creative ideation and evaluation. *ACM: Creativity and Cognition 2011*, 393-394.

Stanovich K. E. (1999). *Who is Rational? Studies of Individual Differences in Reasoning*. Mahwah, NJ: Elrbaum.

Stanovich, K. E. (2004). *The robot's rebellion: Finding meaning in the age of Darwin*. Chicago, IL: University of Chicago Press.

Stanovich, K. E. (2009). Distinguishing the reflective, algorithmic, and autonomous minds: Is it time for a tri-process theory? In J. St. B. T. Evans & K. Frankish (Ed's), *In two minds: Dual processes and beyond*. New York: Oxford University Press. pp. 89–108.

Sternberg, R. J. (1998). Cognitive mechanisms in human creativity: Is variation blind or sighted? *Journal of Creative Behavior, 32,* 159–176.

Vartanian, O., Martindale, C., & Kwiatkowski, J. (2003). Creativity and inductive reasoning: The relationship between divergent thinking and performance on Wason's 2-4-6 task. *Quarterly Journal of Experimental Psychology, 56A*, 641-655.

Vartanian,O.,Martindale,C.,and Kwiatkowski,J.(2007). Creative potential, attention, and speed of information processing. *Personality and Individual Differences*, *43*, 1470–1480.

Vartanian, O., Martindale,C., & Matthews, J. (2009). Divergent thinking ability is related to faster relatedness judgments. *Psychology of Aesthetics, Creativity, and the Arts, 3*, 99-103.

Wallas, G. (1926). *The Art of Thought*. Watts & Co.
Vincent, A. S., Decker, B. P., & Mumford, M. D. (2002). Divergent thinking, intelligence, and expertise: a test of alternative models. *Creativity Research Journal*, *14*, 163-178.
Wallas, G. (1926). *The Art of Thought*. Watts & Co.
Ward, T. B. (1994). Structured imagination: The role of category structure in exemplar generation. *Cognitive Psychology*, *27*, 1–40.
Ward, T. B., & Kolomyts, Y. (2010). Cognition and Creativity in J. C. Kaufman and R. J. Sternberg (Ed's) *The Cambridge Handbook of Creativity*, New York, NY: Cambridge University Press. pp.93-112.
Weisberg, R. W. (2004). On structure in the creative process: A quantitative case-study of the creation of Picasso's Guernica. *Empirical Studies of the Arts, 22,* 23-54.
Zabelina, D. L., & Robinson, M. D. (2010). Creativity as flexible cognitive control. *Psychology of Aesthetics, Creativity, and the Arts, 4*, 136-143.
Zabelina, D. L., & Beeman, M. (2013). Short-term attentional perseveration associated with real-life creative achievement. *Frontiers in Psychology, 4*, 191.

**Table 1.** Comparison of the characteristics of Type 1 and Type 2 processes with dual process models of creative thought. For instance, the table shows that Type 1 processes may contribute to both convergent and divergent thinking in Guilford's (1956) Structure of Intellect model and that Type 2 processes contribute to both generation and exploration of ideas in Finke *et al.'s* (1992) Genoplore model.

| | **Dual process models of cognition (Evans, 2008; Frankish, 2010)** | |
|---|---|---|
| **Dual process model of creativity** | Type 1 processes (rapid, unconscious, automatic, intuitive, associative, unlimited capacity, contextual, reflexive) | Type 2 processes (slow, controlled, effortful, conscious, analytic, capacity limited, rule based, reflective) |
| Cognitive Experiential Self Theory (Epstein, 2003) | Experiential (divergent) thinking | |
| Structure of Intellect (Guilford, 1956) | Divergent thinking<br>Convergent thinking | Divergent thinking<br>Convergent thinking |
| Ideation – Evaluation cycles (Basadur *et al.*, 1982; Basadur, 1985) | Mapping unclear | Mapping unclear |
| Blind Variation and Selective Retention (Campbell, 1960; Simonton, 2011) | Variation (but context effects excluded) | Selection |
| Genoplore (Finke *et al.*, 1992) | Generation | Exploration<br>Generation |
| Dual State Model (Howard-Jones, 2002) | Generative | Analytical |
| Honing Theory (Gabora, 2005) | Associative | Analytic |
| Dual Pathway Model (Nijstad *et al.*, 2010) | | Flexibility<br>Persistence |